\documentclass[iop]{emulateapj}                                                                                                                        

\usepackage{epsfig}
\usepackage{hyperref}
\usepackage{natbib}
\usepackage{graphicx}
\usepackage{amsmath}
\usepackage{multirow}
\usepackage{float}

\usepackage[none]{hyphenat}
\usepackage[export]{adjustbox}

\def\simgt{\lower.5ex\hbox{$\; \buildrel > \over \sim \;$}}
\def\simlt{\lower.5ex\hbox{$\; \buildrel < \over \sim \;$}}

\def\amin{\ifmmode^{\prime}\else$^{\prime}$\fi}
\def\asec{\ifmmode^{\prime\prime}\else$^{\prime\prime}$\fi}

\def\simgt{\lower.5ex\hbox{$\; \buildrel > \over \sim \;$}}
\def\simlt{\lower.5ex\hbox{$\; \buildrel < \over \sim \;$}}

\newcommand\chandra{{\it Chandra}}
\newcommand\Chandra{{\it CHANDRA}}

\newcommand\nustar{{\it NuSTAR\/}}

\def\sgra{Sgr~A$^{\star}$}
\def\SgrA{Sgr~A$^\star$}

\shorttitle{X-ray filament G0.13--0.11}
\shortauthors{S. Zhang et al.}

\begin{document}

\title{\nustar\ and \chandra\ Observation of the Galactic Center Non-thermal X-ray Filament G0.13--0.11: a Pulsar Wind Nebula Driven Magnetic Filament}

\author{Shuo Zhang\altaffilmark{1,2,3}, Zhenlin Zhu\altaffilmark{4,5,6}, Hui Li\altaffilmark{7,8}, Dheeraj Pasham\altaffilmark{7}, Zhiyuan Li\altaffilmark{6}, Ma\"{i}ca Clavel\altaffilmark{9}, \\ Frederick K. Baganoff\altaffilmark{7}, Kerstin Perez\altaffilmark{10}, Kaya Mori\altaffilmark{10}, Charles J. Hailey\altaffilmark{11}}

\altaffiltext{1}{Boston University Institute for Astrophysical Research, 725 Commonwealth Avenue, Boston, MA 02215, USA; shuoz@bu.edu}
\altaffiltext{2}{Bard College Physics Program, 30 Campus Road, Annandale-On-Hudson, NY 12504, USA}
\altaffiltext{3}{NASA Einstein Fellow}
\altaffiltext{4}{SRON Netherlands Institute for Space Research, Sorbonnelaan 2, 3584 CA Utrecht, The Netherland}
\altaffiltext{5}{Leiden Observatory, Leiden University, Niels Bohrweg 2, 2300 RA Leiden, The Netherlands}
\altaffiltext{6}{School of Astronomy and Space Science, Nanjing University, Nanjing 210046, China}
\altaffiltext{7}{MIT Kavli Institute of Astrophysical Research, Cambridge, MA 02139 USA}
\altaffiltext{8}{NASA Hubble Fellow}
\altaffiltext{9}{Institut de Plan\'{e}tologie et d'Astrophysique de Grenoble, Universit\'{e} Grenoble Alpes, CS 40700, Grenoble, France}
\altaffiltext{10}{MIT Department of Physics and Lab of Nuclear Sciences, Cambridge, MA 02139 USA}
\altaffiltext{11}{Columbia Astrophysics Laboratory, Columbia University, New York, NY 10027, USA}

\begin{abstract}
One of the most unique phenomena in the Galactic center region is the existence of numerous long and narrow filamentary structures within a few hundred parsecs of \sgra. 
While more than one than one hundred radio filaments have been revealed by MeerKAT, about two dozens X-ray filaments have been discovered so far.
In this article, we report our analysis on the deep \chandra\ and \nustar\ observations of a non-thermal X-ray filament, G0.13--0.11, which is located adjacent to the Radio arc.
\chandra\ revealed a unique morphology of G0.13--0.11, which is an elongated (0.1~pc in width and 3.2~pc in length) structure slightly bended towards the Radio arc.
A pulsar candidate ($\Gamma\sim1.4$) is detected in the middle of the filament, with a tail of diffuse non-thermal X-ray emission on one side of the filament. 
The filament is detected by \nustar\ up to 79 keV, with the hard X-ray centroid consistent with the pulsar candidate.
We found that the X-ray intensity decays along the filament farther away from the pulsar candidate, dropping to half of its peak value at 2.2 pc away.
This system is mostly likely a Pulsar Wind Nebula interacting with ambient interstellar magnetic field, where the filaments are kinetic jets from PWN as recently proposed.
The nature of this filament adds to complex origin of the X-ray filaments, which serve as powerful tools to probe local and global powerful particle accelerators in the Galactic center.

\end{abstract}
\keywords{X-rays: individual: G0.13--0.11 --- radiation mechanisms:nonthermal}

%%%%%%%%%%%%%%%%%%%%%%%%%%%%%%%%%%%%%%%%%%%%%%%%%%%%%%%%%%%%%

\section{Introduction}

A unique and striking phenomenon in the Galactic center region is the existence of numerous non-thermal radio filaments \citep{Yusef1984, Morris1996, Larosa2000}.
The origin and formation of of these filaments have been long-standing questions for decades.
Dozens of radio filaments as long as tens of parsecs was detected by the Very Large Array within the central $\sim 2^{\circ}$ of the Galaxy, with strong magnetic field ($\sim1$~mG) aligning along the major axis of the filaments \citep{Morris1996, Yusef1988}.
Recent observations of the MeerKAT telescope have revealed more than 100 filaments, which seem to be associated with the newly discovered bi-polar radio bubbles \citep{Heywood2019}.
Thanks to polarization detection, the emission mechanism of non-thermal radio filaments has been pinned down to synchrotron emission.
High-resolution JVLA observations revealed entangled sub-filaments within many radio filaments \citep{Morris2014}.
These results suggest that radio filaments are magnetic structures, where strong and highly-organized magnetic field traps GeV electrons and produces synchrotron emission in the radio band \citep{Zhang2014, Morris2014}.

{Similar filamentary structures, though at smaller spatial scales, have also been detected in the X-ray regime. 
About 30 parsec-long X-ray filaments have been detected so far, some of which have radio counterparts \citep{Muno2008, Lu2008, Johnson2009, Ponti2015}.
A few scenarios have been proposed to explain X-ray filaments:  pulsar wind nebula, supernovae remnant ejecta or shocked molecular cloudlet, magnetic structure like the radio filaments, and ram-pressure or magnetic field confined stellar winds from massive stars.
The combined diagnostic power of broadband X-ray timing, imaging, spectroscopy and multi-wavelength investigation from ratio to TeV bands is the key to distinguish between the above scenarios.

Through \SgrA\ observational campaign and the Galactic center mini survey \citep{Mori2015}, so far \nustar\ has detected a total of four X-ray filaments above 10~keV: G359.95-0.04, G359.97-0.038, G359.89-0.08 (Sgr~A-E), and G0.13--0.11. 
Except for G0.13--0.11, we have revealed the nature of all the other three filaments, two of which were identified thanks to the broadband X-ray morphology and spectroscopy allowed by \nustar.
Our limited sample of hard X-ray filaments shows interesting diversity.
The three identified filaments possess different source nature:
filament G359.95--0.04 is a pulsar wind nebula candidate \citep{Wang2005}; G359.97-0.038 is best explained by a shocked molecular cloudlet adjacent to supernova remnant Sgr A East \citep{Nynka2015, Zhang2018}; G359.89--0.08 (Sgr~A-E) is most likely a magnetic structure fed by TeV electrons \citep{Zhang2014}. 
The remaining question is whether other X-ray filaments would fall into any of the three above scenarios or bring up yet other possibilities.
The nature of the fourth hard X-ray filament, G0.13--0.11, may give us a clue and is thus of vital interest.

Filament G0.13--0.11 borders a bow-shaped radio protrusion from the Radio Arc, which is composed of many semi-aligned radio filaments extending up to a few tens of parsecs \citep{Yusef1984}.
G0.13--0.11 was firstly discussed in \citet{Yusef2002}, and interpreted as part of a diffuse molecular feature emitting 6.4~keV Fe K$\alpha$ line.
Follow-up studies based on \chandra\ data show that G0.13--0.11 demonstrates a non-thermal X-ray spectrum characterized by a featureless power-law model \citep{Wang2005}.
\citet{Wang2005} reported that the filament extends about 40\asec\ (1.6~pc at the distance of 8~kpc) to the southeast from a point source CXOGCS~J174621.5-285256, with an intrinsic width of 2\asec (0.07~pc).
The thin and long filamentary morphology with a slight curvature points to a magnetic-field confined pulsar wind nebula (PWN) as a likely source nature. 

Recently,~H.E.S.S. discovered a new Galactic plane very-high-energy (VHE) point source, HESS~J1746--285 ($l=0.149^{\circ}$, $b=-0.103^{\circ}$), which was confirmed by MAGIC and VERITAS (Archer 2016; Arnen 2016, HESS2017).
This new VHE source is spatially coincident with G0.13--0.11, thus a likely TeV counterpart of the filament.

Recent \nustar\ observations allowed us to study the high-energy X-ray ($>10$~keV) emission from G0.13--0.11 and compare to its soft X-ray and $\gamma$-ray emission.
We also obtained an order of magnitude deeper \chandra\ observation on the G0.13--0.11, which makes it possible to reveal fainter part of this intriguing filament. 
In this paper, we report analysis results based on deep observations of the filament G0.13--0.11 by both \nustar\ and \chandra.
The paper is organized as the following.
In Section 2, we introduce observations used in this study and the data reduction methods.
Next, we present the broadband X-ray morphology of Filament G0.13--0.11 in Section 3, and its X-ray spectral properties in Section 4.
Then we discuss our Spectral Energy Distribution (SED) fitting using multi-wavelength data obtained for this filament.
Finally, in Section 6 we summarize our findings, discuss the nature of this filament and its role in the Galactic center environment.

%%%%%%%%%%%%%%%%%%%%%%%%%%%%%%%%%%%%%%%%%%%%%%%%%%%%%%%%%%%

\section{Observation and Data Reduction}

\subsection{\nustar}

\nustar\ observed the Sgr~A molecular cloud region during the 2012 Galactic center mini-survey, and the Galactic plane survey legacy program conducted in 2016.
Filament G0.13--0.11 is captured in three out of the six pointings ($\sim$25~ks exposure each) of the 2012 mini-survey, and in the 150~ks exposure observation in 2016. 
We used all the four \nustar\ observations to study the high-energy X-ray emission from G0.13--0.11, as listed in Table \ref{tab:obs}.

We reduced the \nustar\ data using HEASOFT v.~6.19, and filtered events for periods of high instrumental background due to South Atlantic Anomaly (SAA) passages and known bad detector pixels. 
Photon arrival times were corrected for on-board clock drift and precessed to the Solar System barycenter using the JPL-DE200 ephemeris.
For each observation, we registered the images with the brightest point sources available in individual observations, improving the astrometry to $\sim 4 \asec$.
For the three 2012 observations (obsID 40010004001, 40010005001, 40010006001), we used the data obtained by both focal plane modules FPMA and FPMB. 
For the 2016 observation (obsID 40202001002), G0.13--0.11 is only captured by FPMA but not FPMB.
We therefore only included FPMA data for the 2016 observation. 

To derive the \nustar\ spectrum of G0.13--0.11, we use a source region with a radius of $r=20\asec$ centered on the source.
Spectra from the same focal plane module are combined and then grouped with a minimum of $3\sigma$ signal-to-noise significance per data bin, except the last bin at the high-energy end for which we require a minimum of $2\sigma$ significant. 

\subsection{\chandra}

The Galactic center region with the filament G0.13--0.11 in the field view was observed by \Chandra\ multiple times during 2000 to 2013. 
We selected two observations obtained in 2000 and 2013 respectively ($\sim50$~ks exposure time each), with G0.13--0.11 at an offset angle smaller than 1\amin\ to avoid PSF distortion at large off-set angles.
The Sgr~A complex observation campaign from 2015 to 2017 (PI: Clavel), including a total of ten observations, contributing to an additional 470.6~ks exposure for this source (Table 1). 
In this dataset, the offset angles for G0.13--0.11 are less than 3\amin. 
These twelve selected \Chandra/ACIS observations add up to a total of ~571.2 ks exposure for G0.13-0.11, reaching one order of magnitude deeper than the observations as used in \citet{Wang2005}.
We processed these observations with CIAO v4.9 and the corresponding calibration files, following the procedure detailed in Zhu et al. (2018). 
We produced a merged event file after astrometric correction using the brightest point sources available in the observations. 
We used all the twelve observations for source and background spectral extraction for the interested regions.

\begin{deluxetable}{lccrr}[H]                                                                                                                
\tablecaption{\nustar\ and \chandra\ observations of G0.13$-$0.11.}
\tablewidth{0pt}
\tablecolumns{4}                                                                                                                    
\tablehead{ \colhead{Instrument} & \colhead{Observation}   &   \colhead{Start Time}   &   \colhead{Exposure} \\
\colhead{ }  & \colhead{ID}  & \colhead{(UTC)}  &  \colhead{(ks)} }  
\startdata
\nustar\  & 40010004001  & 2012-10-15 00:31:07  & 24.0 \\
\nustar\  & 40010005001  & 2012-10-15 13:31:07  & 25.7 \\
\nustar\  & 40010006001  & 2012-10-16 05:41:07  & 23.5 \\
\nustar\  & 40202001002  & 2016-10-28 13:16:08  & 150.9 \\
\hline
\chandra\  &  945       & 2000-07-07 19:04:15      & 49.4 \\
\chandra\  & 14897    & 2013-08-07 16:58:16	 & 51.2 \\
\chandra\  & 17236    & 2015-04-25 14:09:36	 & 80.1 \\
\chandra\  &  17239   & 2015-08-19 17:21:58     &  80.1 \\
\chandra\  &   17237  & 2016-05-18 05:19:05     & 21.1  \\
\chandra\  &  18852   & 2016-05-18 20:49:44     & 53.1 \\
\chandra\  &  17240   & 2016-07-24 05:50:12	 &  75.7 \\
\chandra\  &  17238   & 2017-07-17 18:44:48      &  66.1 \\
\chandra\  &  20118   & 2017-07-23 01:19:57      &  14.1 \\
\chandra\  &  17241   & 2017-10-02 18:53:53      & 25.1 \\
\chandra\  &  20807   & 2017-10-05 13:59:52      &  28.1 \\
\chandra\  &  20808   & 2017-10-08 15:37:12      &  27.1\\
\enddata
\label{tab:obs}
\end{deluxetable}

\begin{figure*}[h!tb]
\centering
\label{fig:morphology}
\begin{tabular}{cc}
\includegraphics[width=0.45\linewidth]{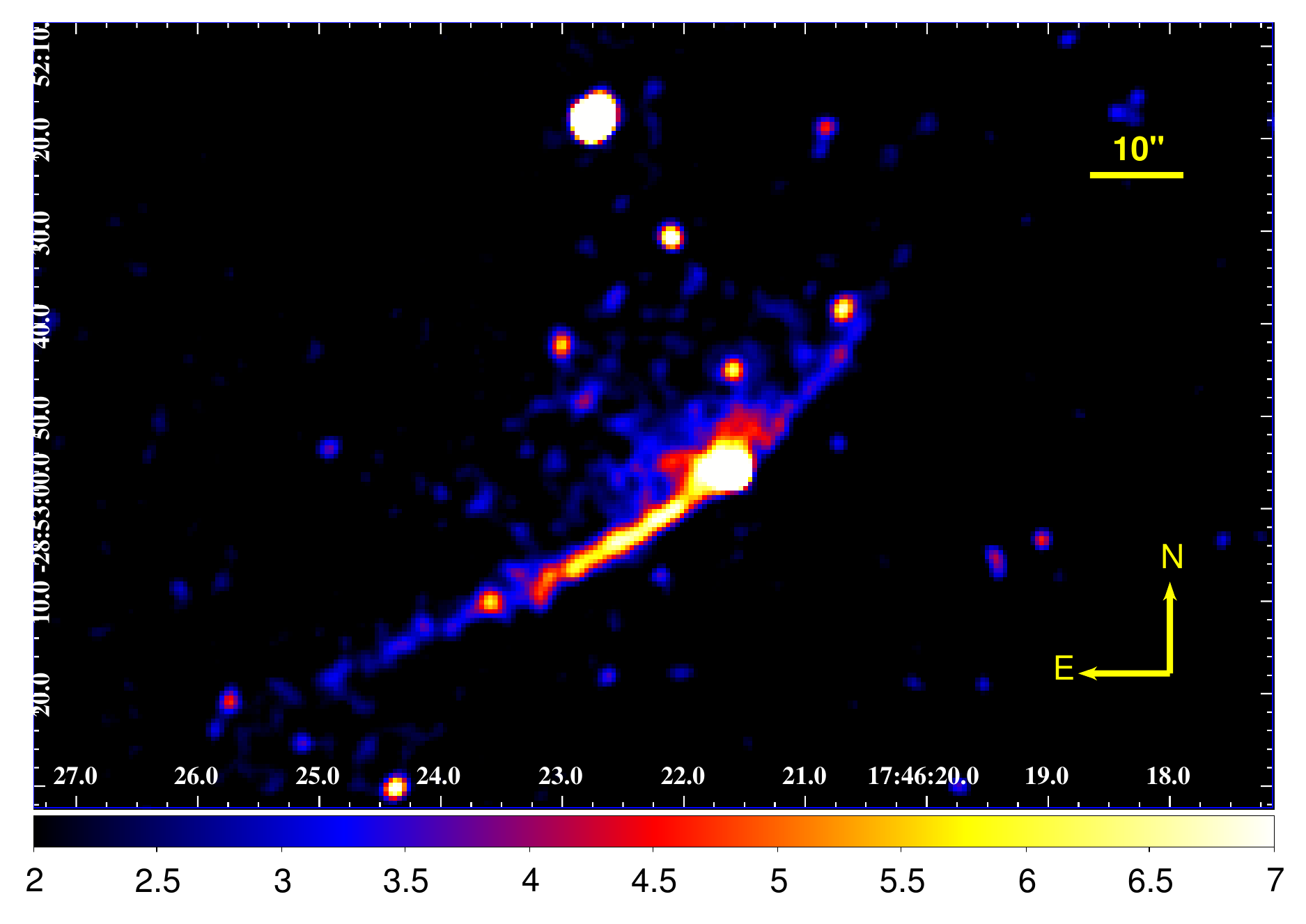} &
\includegraphics[width=0.45\linewidth, angle=0]{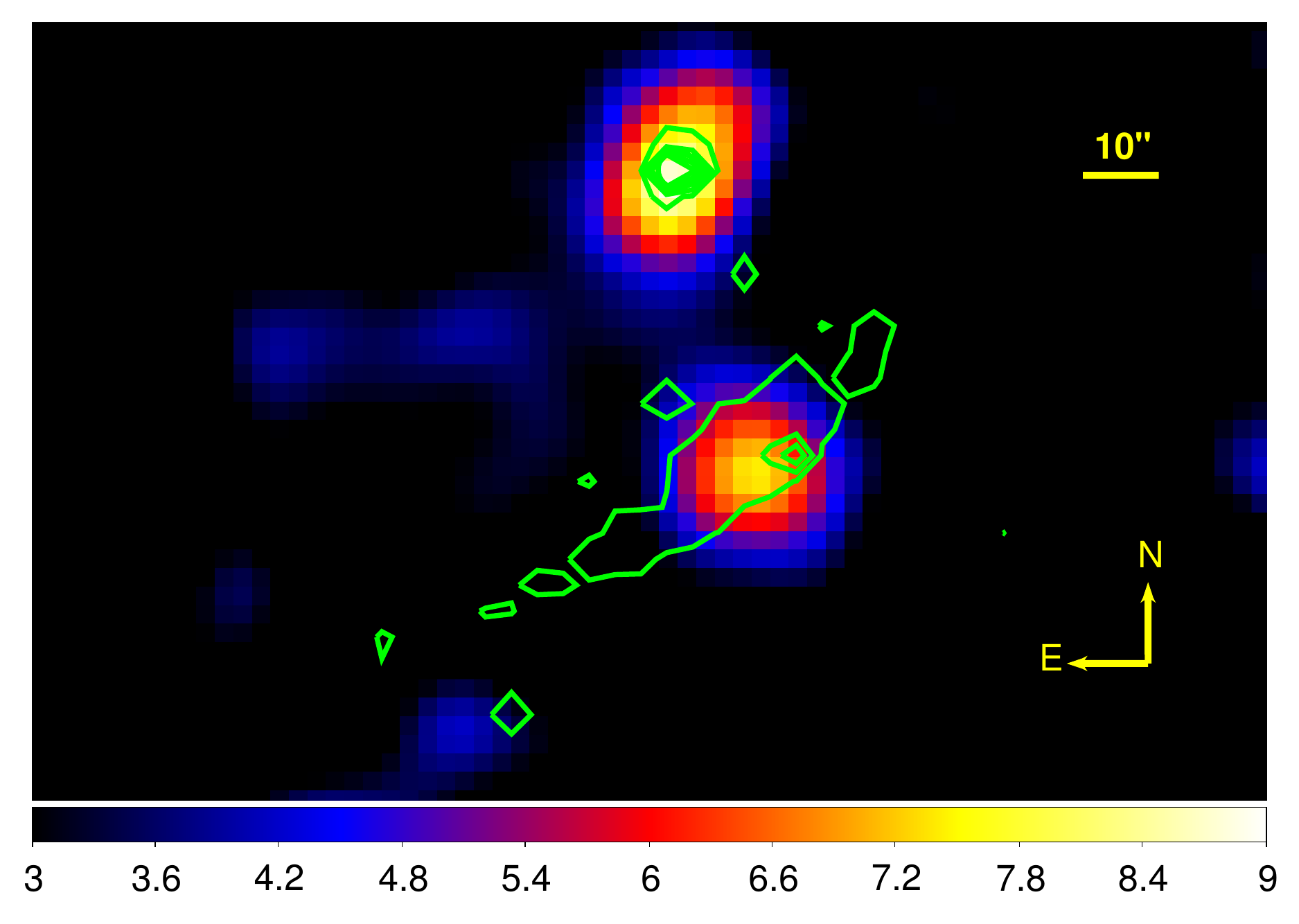}
\end{tabular}
\caption{{\it Left:} \chandra\ 0.5--8~keV~mosaic of the region surrounding Filament G0.13-0.11, showing a thin  (0.1~pc) and long (3.4~pc) structure perpendicular to the Galactic plane, with a slight curvature.
The pulsar candidate CXOGCS~J174621.5--285256 is located at about one third of the filament from the northwest. 
{\it Right:} 10--79~keV \nustar\ mosaic of the Filament region overlaid with green contours made from the 0.5--8~keV \chandra\ image. One point source is detected within the filament, consistent with the position of the pulsar candidate CXOGCS~J174621.5--285256 as detected by \chandra. The rest of the filament is not detectable by \nustar. The other point source in the field is a known intermediate polar CXOUGC J174622.7--285218.}
\end{figure*}  

%%%%%%%%%%%%%%%%%%%%%%%%%%%%%%%%%%%%%%%%%%%%%%%%%%%%%

\section{Broadband X-ray Morphology of Filament}

The left panel of Figure 1 shows the 0.5--8~keV mosaic image made from of all the 12 \chandra\ observations (see Table 1) in the region around the filament G0.13--0.11.
This filament is located at about 30~pc from \sgra\ on the projected plane (assuming the Galactic center at a distance of 8~kpc), in a region occupied by the giant molecular cloud G0.11--0.11.
The orientation of this filament is roughly perpendicular to the Galactic plane.

Deep \chandra\ observations allow us to resolve fainter parts of the filament, giving a fuller picture of its morphology.
The entire filament is $\sim86$\asec\ ($\sim3.4$~pcs) in length and $\sim3$\asec\ ($\sim0.1$~pc) in width (left panel of Figure 1), twice the length reported in \citet{Wang2005} ($\sim40$\asec\ in length and $\sim2$\asec\ using about 100~ks data).
The point source CXOGCS~J174621.5--285256 is detected at R.A.=17$^{h}46^{m}21.62^{s}$, Decl.=--28$^{\circ}$52\amin56.2\asec, located at about one third of the filament from the northwest.  
There is another bright point source detected nearby, which is CXOUGC J174622.7--285218 located at R.A.=17$^{h}46^{m}22.73^{s}$, Decl.=--28$^{\circ}$52\amin22.7\asec.
It is a intermediate polar (IP) with a high temperature of $\rm kT\sim30$~keV, as reported in \citet{Hong2016}.

The filament demonstrates a slight curvature, bending towards the northeast, i.e.~the direction pointing to the Radio Arc and away from \sgra.
The southwest side of the filament shows a sharp edge,
while on the opposite side of the filament there is a fuzzy region with an excess of X-ray emission around the pulsar candidate.
Such morphology suggests a picture where a pulsar, with pulsar wind nebula tail, is moving towards southwest, while hitting into a highly organized magnetic structure and illuminating it with extremely relativistic electrons accelerated by the pulsar.

The right panel of Figure 1 shows the 10--79~keV mosaics of \nustar\ observation of the region around G0.13--0.11.
At the position of the filament, a point-like source coinciding with the \Chandra\ point source is detected up to 79~keV. 
We overlaid the high-energy X-ray mosaic with green contours made from the 0.5--8~keV \chandra\ image.
The centroid of the point-like \nustar\ source is $\sim4$\asec\ from the Chandra position of the pulsar candidate CXOGCS~J174621.5--285256, within the \nustar\ position uncertainty after astrometric correction.
Therefore, the point-like X-ray source \nustar\ detected is mostly the hard X-ray counterpart of the pulsar candidate CXOGCS~J174621.5--285256.
The extended emission along the filament is not resolved by \nustar\ with the current dataset.
The nearby IP CXOUGC J174622.7--285218 is also significantly detected beyond 10~keV, due to its high temperature.

%%%%%%%%%%%%%%%%%%%%%%%%%%%%%%%%%%%%%%%%%%%%%%%%%%%%%%%%%%%%%
\section{X-ray Spectra and Intensity Profile of the Filament}

\subsection{Spectrum of the Whole Filament}

For the \chandra\ data, we firstly extracted the source spectrum of the entire filament from each individual \chandra\ observation, and the background spectrum from the selected sky region (Figure 3).
We then fit the background-subtracted spectrum for each observation using a  simple absorbed power-law model {\tt tbabs*powerlaw}, which can best fit all the spectra.
We found that the best-fit model parameters for all the spectra are consistent among observations, suggesting no flux variability or spectral evolution for this filament from 2000 to 2017.
Therefore, next we added up all the individual source spectrum, added up all the background spectrum, to derive a combined spectrum set for the filament for the following analysis.
For the \nustar\ data, we extracted the source spectrum from a circular region centered on the point-like source for each individual observation.
We carefully selected a partial ring region, which is in the same sky region as the \chandra\  background region  (Figure 3), at the same distance of the \nustar\ point source from the nearby IP, to compensate for the PSF wing contamination from the IP CXOUGC J174622.7--285218.
After confirming no significant source or background variability among observations, we then combined all the \nustar spectra as well for the following analysis.

We then jointly fit the combined background subtracted \chandra\ spectrum in 2--8~keV and the \nustar\ spectrum in 3--79~keV. 
While the fainter part of the filament falls below the threshold of the \nustar\, while \chandra\ detected the full signal from the filament including the point source, \nustar\ can only detect a partial signal from the filament which is above the threshold, i.e. the point source and the brightest sections of the filaments which are not resolvable.
Therefore, to evaluate this difference between \chandra\ and \nustar\ detection we introduced a constant into the models to evaluate which fraction of the whole \chandra\ source signal can NuSATR detect, resulting a model of  {\tt const*tbabs*powerlaw}.
We set the constant parameter for the \chandra\ model as 1, and the constant for the \nustar\ model as free. 
All the other parameters for the two spectra sets are linked.
The spectrum is featureless and can be best fit by a simple absorbed power-law model in 2--79~keV.
The data does not require a spectral break or cutoff up to 79~keV, confirming a non-thermal nature.
The best-fit photon index is $\Gamma=1.7\pm0.2$ (error bars at 90\% confidence level, same for the following).
The absorption column density results in $N_{\rm H}=(1.2\pm0.2)\times10^{23}~\rm cm^{-2}$ ($\chi^2_{\rm \nu}=0.9$ for $d.o.f.=207$).
The constant for the \nustar\ model is found to be $0.55\pm.0.08$, i.e. $\sim55$\% of the entire filament is detectable by \nustar.
The observed flux of the entire filament as detected by \chandra\ is $F_{2-10}=(3.1\pm0.1)\times10^{-13}\rm~erg~cm^{-2}~s^{-1}$ in 2--10~keV and $F_{10-79}=(1.1\pm0.2)\times10^{-12}\rm~erg~cm^{-2}~s^{-1}$ in 10--79~keV, corresponding to luminosities of  $L_{2-10}=(3.8\pm0.2)\times10^{33}\rm~erg~s^{-1}$  in 2--10 keV and $L_{10-79}=(9.2\pm0.2)\times10^{33}\rm~erg~s^{-1}$ in 10--79~keV, assuming a distance of 8~kpc.

\begin{figure}[h!tb]
\centering
\label{fig:morphology}
\includegraphics[width=0.6\linewidth, angle=270]{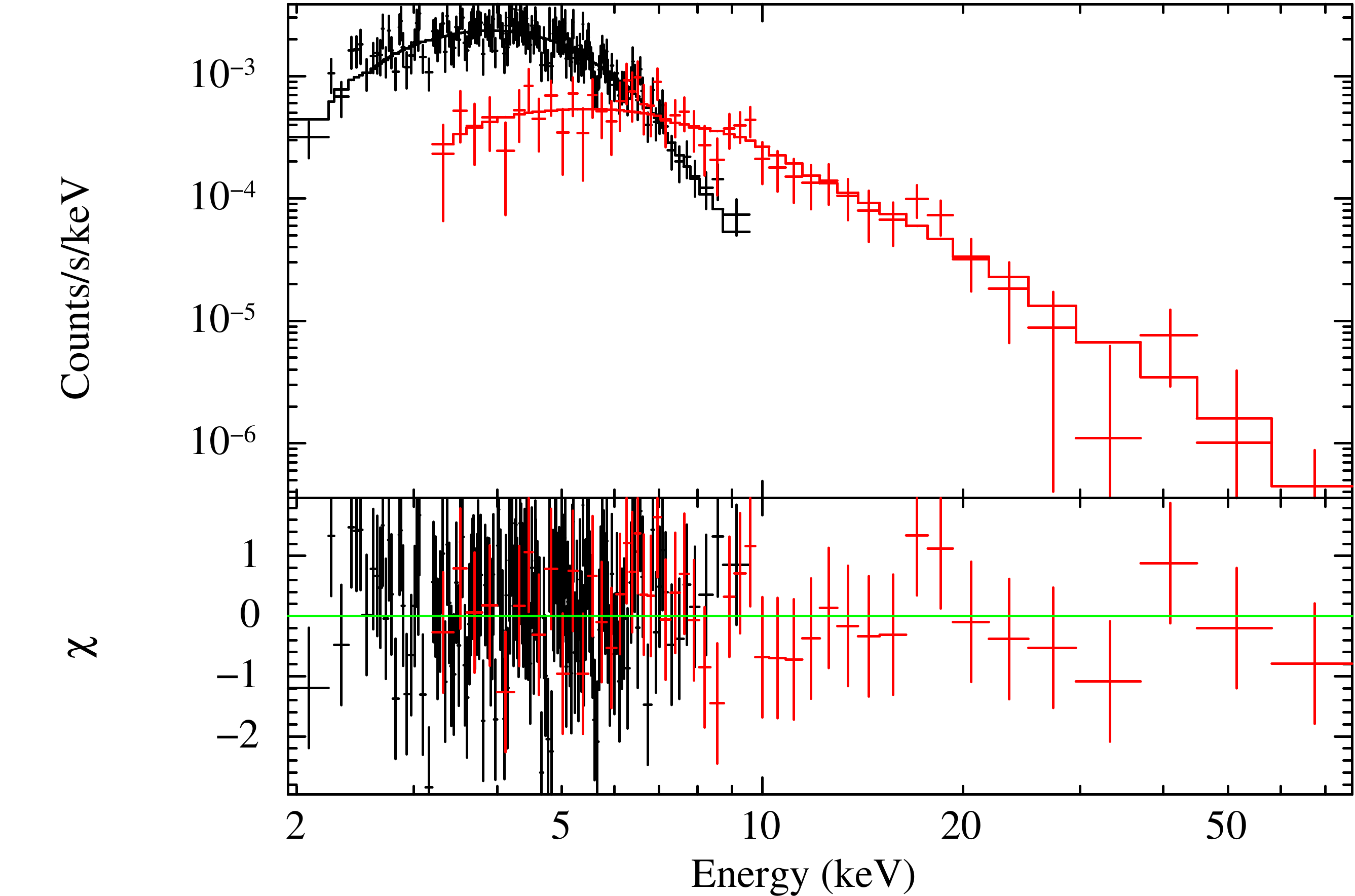} 
\caption{Joint spectral fit for 2--8~keV \chandra\ spectrum and 3--79~keV \nustar\ spectrum for the entire filament. The spectra can be fit an absorbed power-law model, with an absorption column density of $N_{\rm H}=(1.2\pm0.2)\times10^{23}~\rm cm^{-2}$  and a photon index of $\Gamma=1.7\pm0.2$. No spectral break of cutoff is required by the data.}
\end{figure}

\begin{figure*}[t]
\centering
\label{fig:spec3parts}
\begin{tabular}{rr}
\includegraphics[width=0.38\linewidth, angle=0]{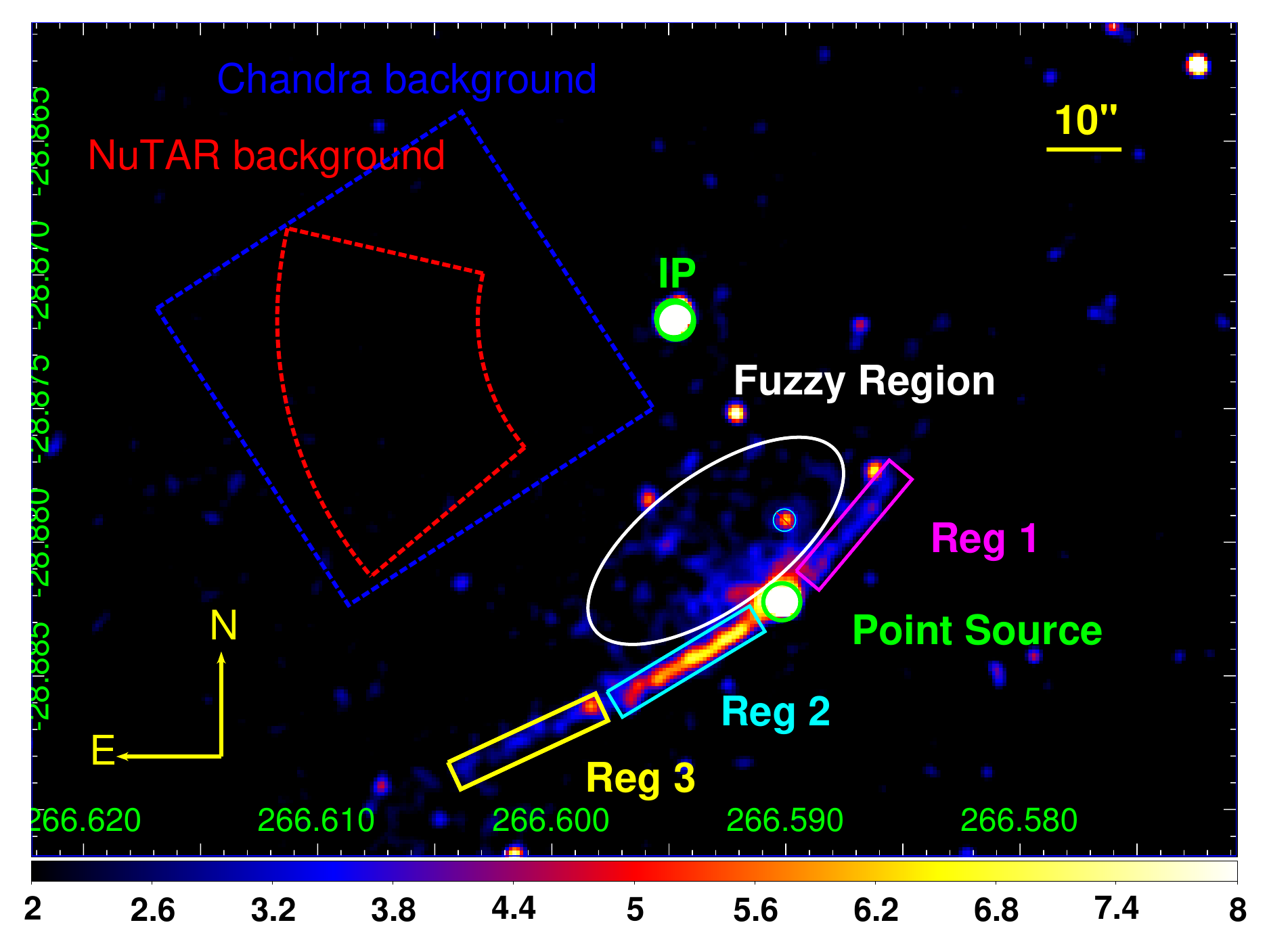} &
\includegraphics[width=0.39\linewidth, angle=0]{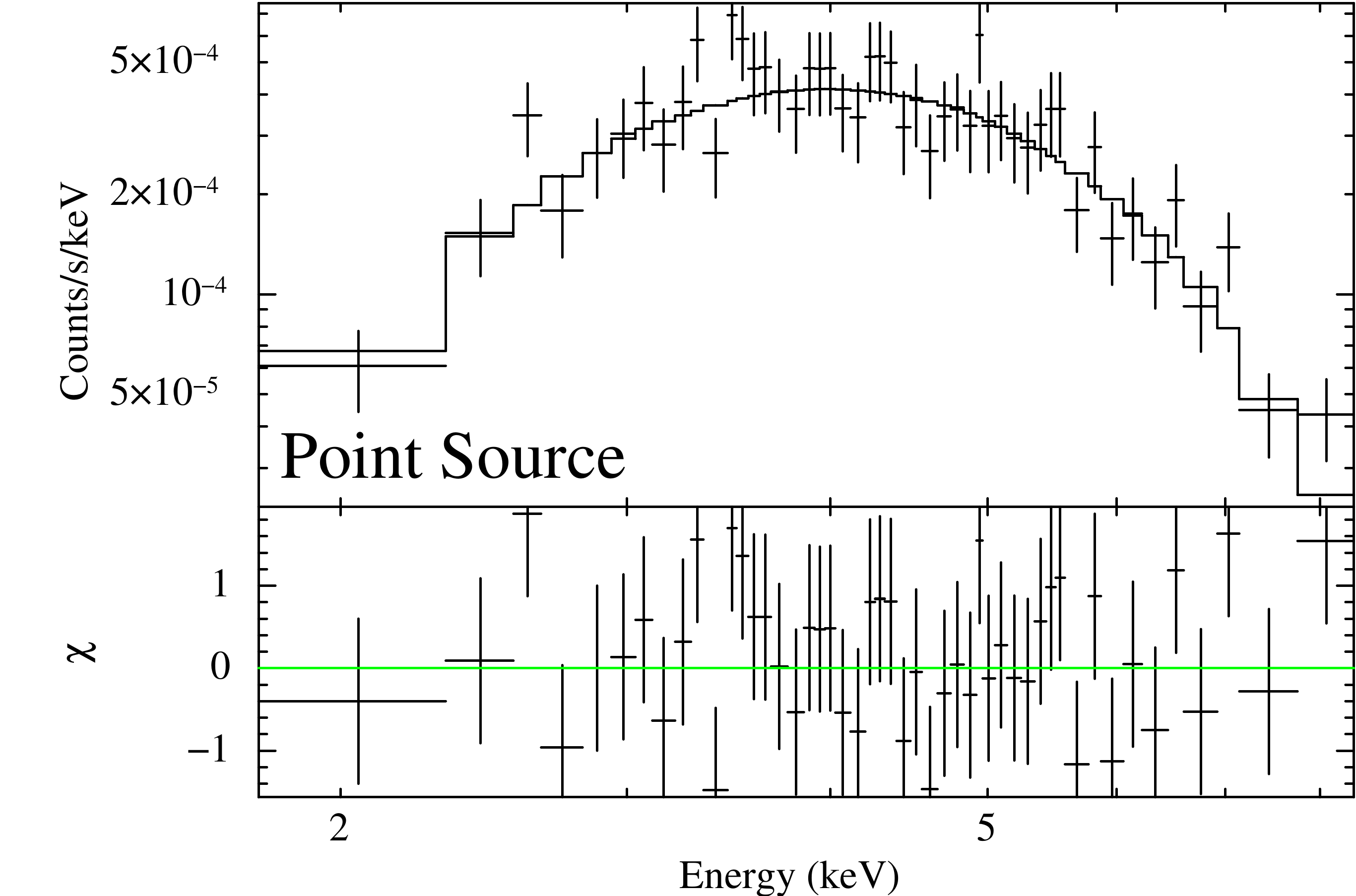} \\
\includegraphics[width=0.4\linewidth, angle=0]{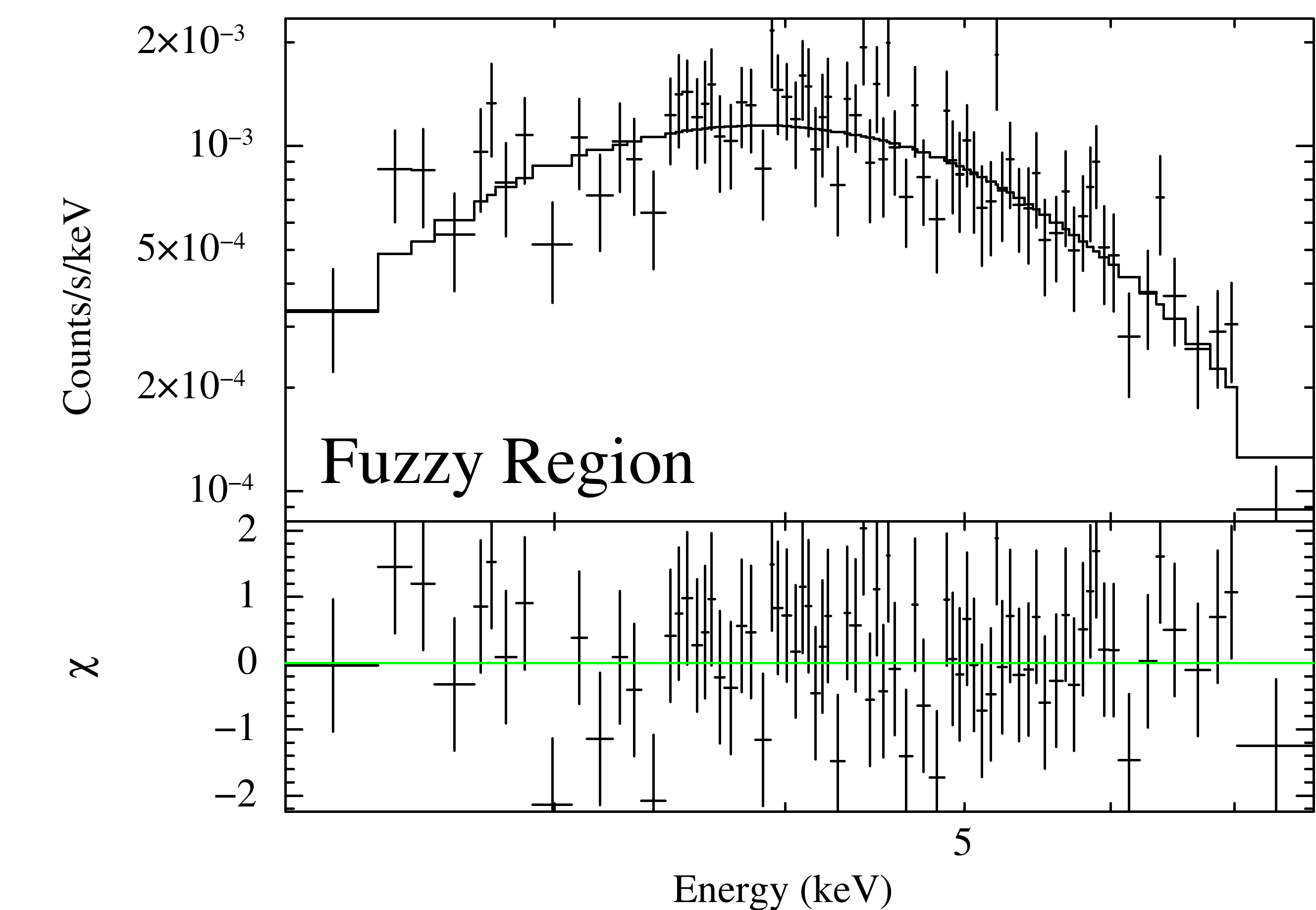} &
\includegraphics[width=0.4\linewidth, angle=0]{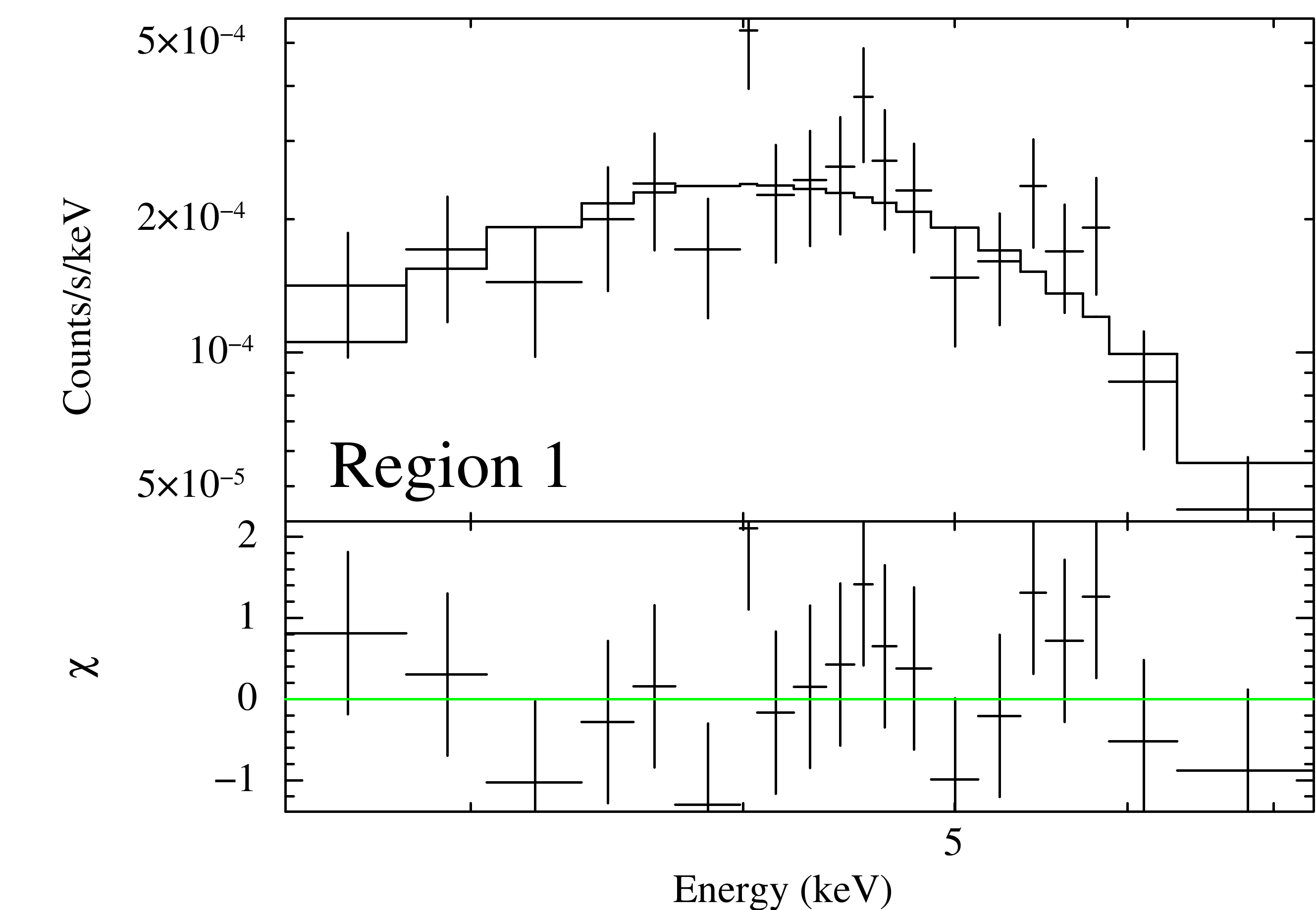} \\
\includegraphics[width=0.4\linewidth, angle=0]{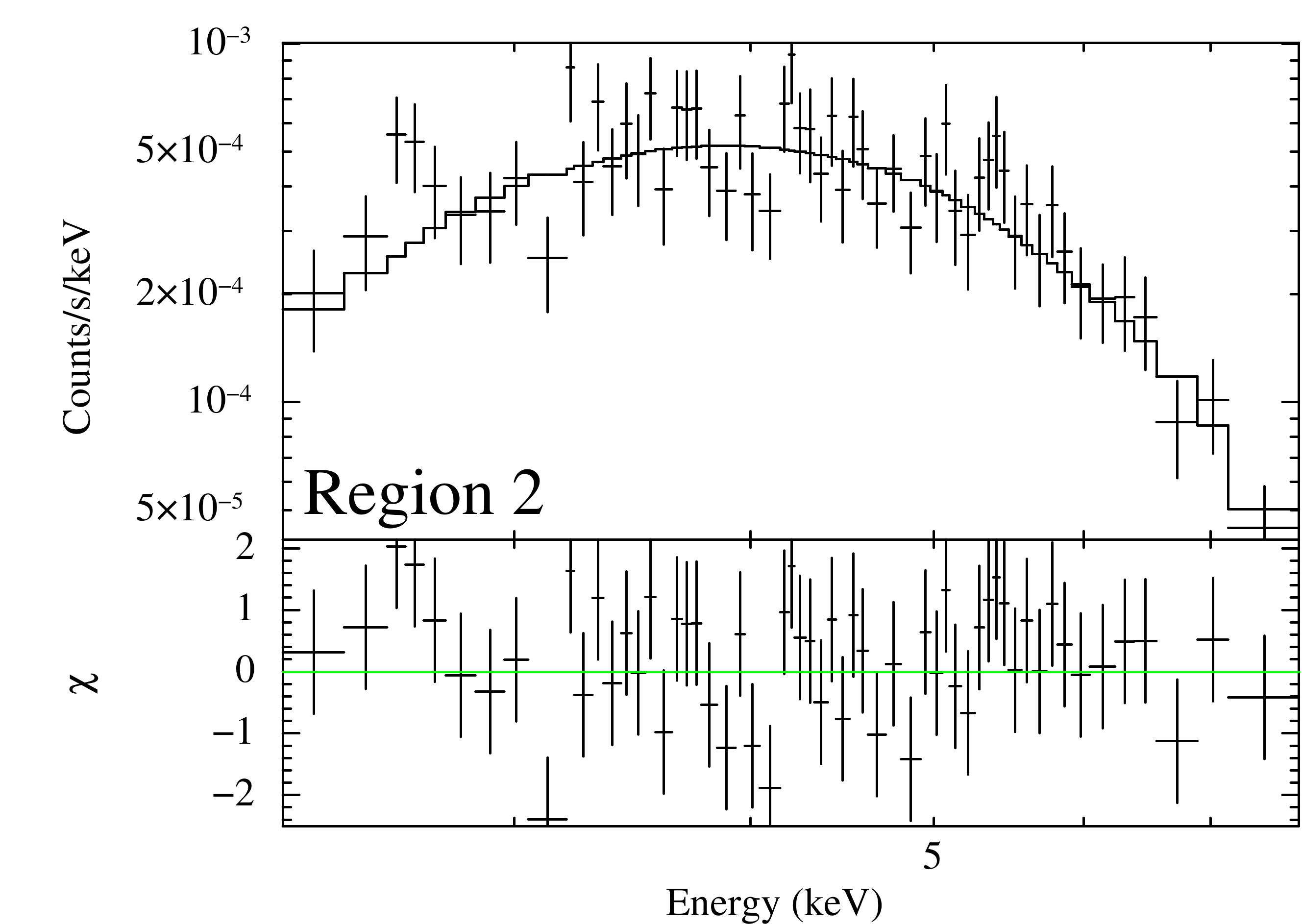} &
\includegraphics[width=0.4\linewidth, angle=0]{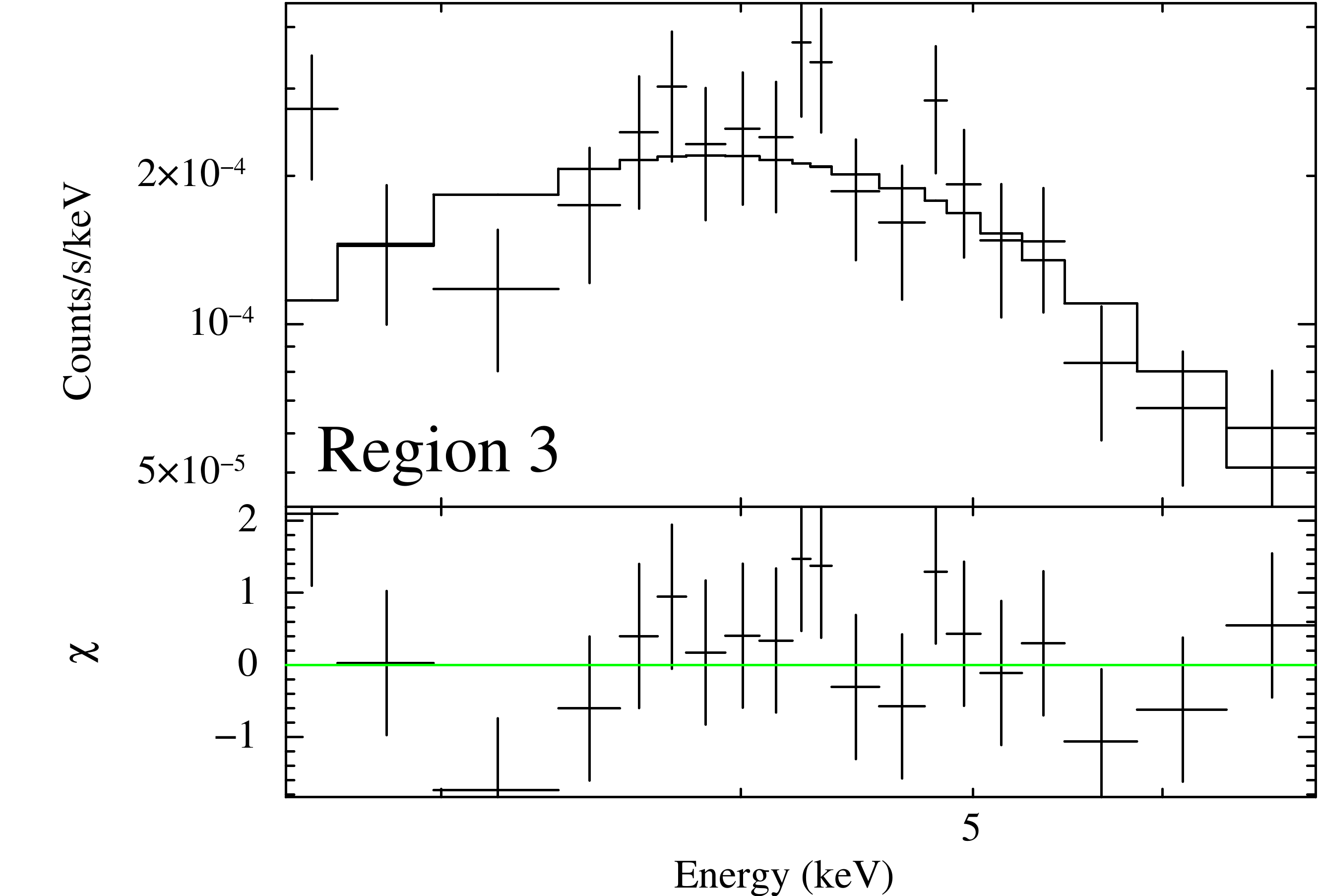} \\
\end{tabular}
\caption{Spatially resolved spectra for five different regions within and around the filament. Upper left: 2--8 keV image of the filament showing the five regions from which spectra were extracted: a $r=2.5$\asec circle region centered on the pulsar candidate (green circle); an elliptical region on the northeast side of the filament, one point source within it excluded (white ellipse); and three 4\asec $\times$ 25\asec\ rectangular regions along the filament, dubbed as Reg 1, Reg 2 and Reg 3. Corresponding spectra extracted from these five selected regions are shown in the other panels. The \Chandra\ spectra from all the five regions can be well fit to absorbed power-law models. The best-fit photon indices for different regions are listed in Table 2.}
\end{figure*}

\begin{deluxetable*}{llcccc}[h!tb]                                                                                                         
\tablecaption{Spatially Resolved Spectral Information.}
\tablewidth{0pt}
\tablecolumns{6}                                                                                                                  
\tablehead{ \colhead{Source} & \colhead{Net Source Counts} & \colhead{$N_{\rm H}$}   &   \colhead{$\Gamma$}   &   \colhead{$F_{obs(2-10\rm keV)}$}  & \colhead{$L_{2-10\rm keV}$} \\
\colhead{ }  & \colhead{ } & \colhead{($10^{23}~\rm cm^{-2}$)}  & \colhead{ }  &  \colhead{(\rm $10^{-14}\rm~erg~cm^{-2}~s$)}  & \colhead{($\rm 10^{32}~erg~s^{-1}$)}}  
\startdata
Point Source		& $825\pm32$ 			& $1.2\pm0.3$ 		& $1.4\pm0.5$  	& $7.2\pm0.4$	        & $8.4\pm0.8$   \\
Fuzzy Region	 	& $2124\pm74$ 	& $1.1\pm0.4$  		& $1.7\pm0.5$ 		& $15.1\pm0.7$ 	& $18.7\pm1.2$      \\
Region 1		  	& $430\pm28$  	& $1.1\pm0.9$	 	& $1.6\pm1.1$ 		& $3.5\pm0.4$	        & $3.9\pm0.5$    \\
Region 2   		& $978\pm36$   	& $1.2\pm0.3$          & $1.8\pm0.3$ 		& $6.7\pm0.5$          & $8.6\pm0.6$     \\
Region 3   		& $377\pm25$  	& $1.3\pm0.8$  	& $1.9\pm1.1$ 		& $3.0\pm0.3$	  	& $ 3.7\pm0.1$   
\enddata
\footnote{The source regions are a $r=2.5\asec$ circle centered on the pulsar candidate for the point source,  an elliptical region to the northeast of the pulsar candidate for the fuzzy region, and $4\asec\times$25\asec\ rectangular boxes for Regions~1-3, as shown in the upper left panel of Figure 2.}
\label{tab:spec}
\end{deluxetable*}

\subsection{Spatially Resolved Spectra and Intensity Profile}
  
To investigate whether there is any spectral evolution along the filament, we studied spatially resolved spectra using all the available \chandra\ data. 
We extracted spectra from five different regions within and close to the filament: a $r=2.5\asec$ circular region centered on the pulsar candidate; the fuzzy region with excessive X-ray emission around the pulsar on the northeast side of the filament; and three $4\asec\times25\asec$ rectangular regions along the filament, with Region 1 on the northwest side of the pulsar candidate and Region 2\&3 on the southeast side (see the upper left panel of Figure 2).

The \chandra\ spectra from all the five regions can be well fit to absorbed power-law models.
The best-fit photon indices are $\Gamma=1.4\pm0.5$ for the pulsar candidate, $\Gamma=1.6\pm1.1$ for Region 1, $\Gamma=1.8\pm0.3$ for Region 2, $\Gamma=1.7\pm1.1$ for Region 3.
Although the best-fit values of photon indices for the three regions ($\Gamma\sim1.5-1.8$) along the filament are larger than the pulsar ($\Gamma\sim1.4$), spectral softening away from the point source is not significant given the error bars. 
The fuzzy region to the northeast of  the point source turns out to also have a featureless power-law spectrum with $\Gamma=1.7\pm0.5$, which is softer than the point source and typical for a pulsar wind nebula.
Therefore, this fuzzy region is highly likely from the PWN behind the pulsar moving towards southwest. 
Figure  3 shows spectra for all the five regions discussed above.

In order to  derive the filament surface brightness as a function of distance to the pulsar candidate, we further divided the southern branch of the filament into eight sections along the major axis and calculated photon flux for each section.
The resultant 2--8 keV intensity profile is shown in Figure 3. 
The emission intensity roughly linearly decays as it gets further away from the pulsar candidate along the filament, dropping to 50\% of its peak intensity at $\sim51$\asec\ (2.2~pc) from the point source.
In conclusion, although significant spectral softening away from the pulsar candidate is not detected, the X-ray emission intensity drops along the filament.

\begin{figure}
\centering
\label{intensity}
\includegraphics[width=0.9\linewidth, angle=0]{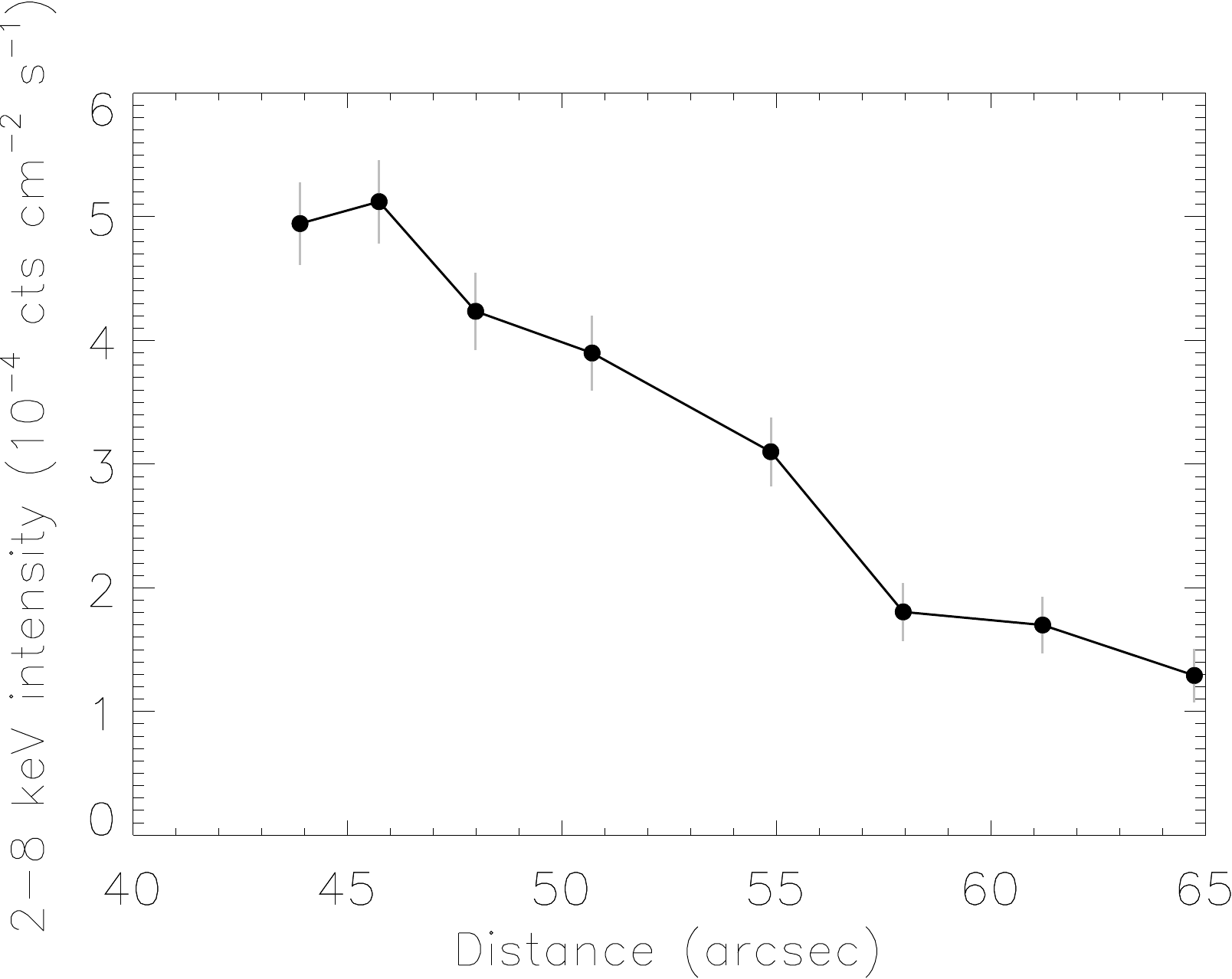}
\caption{2-8 keV Intensity profile along the filament. The error bars are at 1-$\sigma$ confidence level. The X-ray emission intensity along the filament decays as it gets further away from the pulsar candidate, dropping 50\% at 2.2~pc from the pulsar.}
\end{figure}

\section{Summary and Discussion}

\subsection{Origin of the Filament G0.13-0.11}
The filament G0.13--0.11 is located at 30~pc from \sgra, and right next to the Radio Arc.
A bow-shaped radio protrusion from the radio arcs extends to the position of Filament G0.13--0.11, while G0.13--0.11 itself does not have a radio counterpart.
G0.13--0.11 processes a unique X-ray morphology, with a thin and long linear structure slightly bended towards the Radio Arc.
With one order of magnitude deeper observation (from 50~ks to 571~ks), \chandra\ revealed fainter part of the Filament G0.13--0.11 compared to previous work \citep{Wang2005}.
The whole filament is now detected as $\sim3.2$~pc in length and $\sim0.1$~pc in width. 
Based on the length of the southern branch of the filament ($\sim2$~pc) and the synchrotron lifetime of the X-ray emitting particles, we confirm that the magnetic field strength along the filament is $B\le0.3$mG as previously estimated in \citet{Wang2005}.
With the given magnetic field strength, it requires injection of TeV electrons to produce X-ray emission up to a few tens of keV as detected by \nustar.

A bright non-thermal X-ray point source ($\Gamma=1.4\pm0.5$) resides in the middle of the filament, with a tail of non-thermal emission ($\Gamma=1.7\pm0.5$) on the northeast side of the filament, the direction towards the Radio Arc. 
High-energy X-ray emission up to 79~keV is detected by \nustar\ from G.13-0.11, with its centroid consistent with the location of the point source.
We found that the intensity roughly linearly decays away from the point source, dropping to half of its peak intensity at $\sim51$\asec\ (2.2~pc) from the point source.
Furthermore, spatially resolved spectral analysis along one branch of the filament shows a trend of spectral softening along the southern branch of the filament as it gets farther away from the point source, though not significant given the error bars.
If this spectral softening trend can be confirmed by future data, it will further support the scenario that there is a local particle accelerator within this filament.
And the reason why the northern branch of the filament is significantly fainter than the southern branch could be due to the Doppler effect, which implies further suggests substantial flow velocity along the filament.

Considering the non-thermal X-ray spectrum with a photon index of $\Gamma=1.4\pm0.5$, and an X-ray luminosity of $L_{X}=(8.4\pm0.8)\times10^{32}$~erg~s$^{-1}$, the point source within the filament (CXOGCS~J174621.5--285256) is most likely a young pulsar.
We therefore performed a pulsation search for a $r=20$\asec\ region centered on the pulsar candidate in different energy bands using the \nustar\ data. 
A pulsation signal is not detected at a 99\% confidence level with the current data.
Additional exposure by the \nustar\ telescope is essential for detection of X-ray pulsation signals using a wide energy band. 
Non-thermal X-ray emission from the pulsar can be produced by magnetospheric emission originating from the co-rotating magnetosphere.
The fuzzy region on the northeast side of the filament demonstrates excessive extended non-thermal emission ($\Gamma=1.7\pm0.5$).
This extended emission mostly likely originates from a pulsar driven synchrotron nebula, which suggests that the pulsar is moving towards the southwest, thus leaving its nebula tail behind.

The emission along filament is best explained by magnetic structures fed by relativistic particles produced by the young pulsar.
One possibility proposed recently for such thin and long feature from a point source is kinetic jets or misaligned outflow from ram-pressure confined PWN\citep{Barkov2019a, Barkov2019b}, as seen from the Lighthouse PWN, the Guitar PWN and similar systems which also demonstrate remarkably linear structures connected to a PWN \citep{Bandiera2008, Pavan2016}.
In the case of G0.13--0.11, the PWN gets confined by the magnetic field to one side of the magnetic structure, resulting excessive non-thermal X-ray emission in the fuzzy region on the northeast side of the filament but not the other. 
The long and narrow features are produced through reconnection between bow-shock PWN and ambient interstellar medium \citep{Bykov2017, Olmi2019}. 

The highly-organized magnetic field structures can be pre-existing, reflecting turbulent magnetic field caused by energetic activities in the Galactic center region. 
When a pulsar runs into such a magnetic field structure, the escaped TeV electrons/positrons accelerated by the pulsar magnetosphere can be confined in the magnetic field structure, light it up and form the jet-like features.
We note that there are X-ray filaments with no obvious local particle accelerator, which calls for other mechanism for the production of TeV electrons, as discussed in the next session.

\subsection{Origin of Galactic Center Non-thermal Filaments}
The nature of the radio filaments is generally better known: magnetic structures illuminated by GeV electrons.
Based on recent MeerKAT results, the GeV electrons mostly likely come from a bursting event in the vicinity of \sgra, which drives a bipolar outflow and accelerated electrons at the shock front \citep{Heywood2019}, though some radio filaments might be powered by local particle accelerators such as PWNe and SNRs \citep{Barkov2019a, Barkov2019b}.
However, the X-ray filaments seem to show more complex origins.
So far, we have thoroughly studied four X-ray filaments with broadband X-ray data obtained by \nustar\ and \chandra, which lead us to conclude that: one of them is a PWN; a second is a cloudlet lit up by SNR shock; a third is a magnetic structure lit up by $\sim100$~TeV electrons, and the fourth, i.e. G0.13--0.11 discussed in this work, is kinetic jets from a PWN.
The diversity in the origin of the X-ray filaments calls for a protocol to effectively distinguish between PWN, SNR shock-cloud interaction, magnetic structure, or a combination of them.
Head-tail morphology, pulsation detection, and spectral softening are evidence for PWN; proximity to an SNR, SNR-cloud interaction signature like OH 1720 MHz maser and SNR ejecta emission lines would be characteristic for SNR-cloud interaction; while resolved fine-structure of linear structure or even sub-filaments, polarization detection demonstrating magnetic field lines along the filament, would strongly favor the magnetic structure scenario.
The remaining questions are: Is there a dominating mechanism for X-ray filaments? What is the relationship between X-ray and radio filaments? Are X-ray magnetic filaments of the same origin with the radio filaments?
The combined diagnostic power of X-ray morphology, spectrum, timing, and a systematic comparison with high-resolution radio data like those obtained from MeerKAT, would be essential to address these questions.

Among the above proposed filament nature, the magnetic structure scenario is the most intriguing, which can explain the radio filaments and some X-ray filaments we have found.
When TeV electrons/positions enter the highly-organized magnetic structures, possibly with locally enhanced magnetic field, they would produce synchrotron emission falling into the X-ray band; while GeV electrons would produce radio synchrotron emission. 
But the ultimate questions is: where does the GeV and even TeV electrons come from?
In the case discussed in this work, the filament G0.13--0.11 contains an obvious local particle accelerator within it, i.e. a pulsar, which could accelerate electrons to the required energies. 
However, other X-ray filaments like Sgr A-E does not contain a local accelerator within or in proximity to the filament \citep{Zhang2014}. 

Below we propose a mechanism for the production of TeV electrons that can feed the X-ray non-thermal filaments within no local particle accelerator.
Recent TeV $\gamma$-ray observations of the Galactic center show a strong correlation between TeV emission and molecular gas clouds, pointing to a hadronic process, where PeV protons collide into ambient gas and that the proton-proton interaction
produces numerous secondary particles, including neutral pions which decay into $\gamma$-ray photons \citep{HESS2016}.
Distribution of $\gamma$-ray emission along the Galactic plane suggests that the PeV proton accelerator (or a PeVatron) exists within the central 10 parsecs of the Galaxy, pointing to \sgra\ as the PeVatron candidate. 
Another secondary particles of collision of PeV protons within the molecular clouds are relativistic electrons.
According to \citet{Gabici2009}, secondary electrons with energies ranging from GeV to a few hundred TeV can escape from the molecular clouds before cooling off.
If the highly-organized magnetic structures are close enough, or attached to the molecular clouds, they can capture the escaped $\sim100$~TeV electrons, which will produce synchrotron X-ray emission within the filaments. 
Actually, the majority of the filaments detected so far are indeed overlapping with gas clouds. 
The brightest X-ray filament in the Galactic center, Sgr A-E, produces X-ray emission through such a mechanism. 
Since there is no efficient TeV electron accelerator nearby, TeV electrons feeding Sgr A-E are most likely secondary particles from hadronic process of PeV protons. 
Therefore, this sub-set of X-ray filaments could provide an independent approach to indirectly probe PeV protons and their accelerator. 
To test this hypothesis, first necessary steps are a thorough mapping of X-ray filaments, and discovery of more filaments using the next-generation X-ray telescope with high-throughput, high spatial resolution, and polarization detection, like Athena, IXPE and eXTP.

\acknowledgements
This work made use of data from the \nustar\ and \chandra\ missions. We thank both the \nustar\ and \chandra\
Operations, Software and Calibration teams for support with the execution and analysis of these observations. 
S.Z. acknowledges support from NASA through the NASA Hubble Fellowship grant
\# HST-HF2-51450.001-A awarded by the Space Telescope Science Institute, which is operated by the
Association of Universities for Research in Astronomy, Inc., for NASA.
M.C. acknowledges financial support from the French National Research Agency in the framework of the ``Investissements d'avenir" program (ANR-15-IDEX-02) and from CNES.
H.L. is supported by NASA through the NASA Hubble Fellowship grant HST-HF2-51438.001-A awarded by the Space Telescope Science Institute, which is operated by the Association of Universities for Research in Astronomy, Incorporated, under NASA contract NAS5-26555.

\end{document}